# AmorProt: Amino Acid Molecular Fingerprints Repurposing-based Protein Fingerprint

Myeonghun Lee[1,*] and Kyoungmin Min[2,*]

[1]School of Systems Biomedical Science, [2]School of Mechanical Engineering, Soongsil University, 369 Sangdo-ro, Dongjak-gu, Seoul 06978, Republic of Korea

## ABSTRACT

As protein therapeutics play an important role in almost all medical fields, numerous studies have been conducted on proteins using artificial intelligence. Artificial intelligence has enabled data-driven predictions without the need for expensive experiments. Nevertheless, unlike the various molecular fingerprint algorithms that have been developed, protein fingerprint algorithms have rarely been studied. In this study, we proposed the amino acid molecular fingerprints repurposing-based protein (AmorProt) fingerprint, a protein sequence representation method that effectively uses the molecular fingerprints corresponding to 20 amino acids. Subsequently, the performances of the tree-based machine learning and artificial neural network models were compared using (1) amyloid classification and (2) isoelectric point regression. Finally, the applicability and advantages of the developed platform were demonstrated through a case study and the following experiments: (3) comparison of dataset dependence with feature-based methods; (4) feature importance analysis; and (5) protein space analysis. Consequently, the significantly improved model performance and data-set-independent versatility of the AmorProt fingerprint were verified. The results revealed that the current protein representation method can be applied to various fields related to proteins, such as predicting their fundamental properties or interaction with ligands.

*Corresponding author; M. Lee (leemh216@gmail.com), K. Min (kmin.min@ssu.ac.kr)

**INTRODUCTION**

Advances in artificial intelligence have had a significant impact on drug discovery, providing opportunities to accelerate and streamline drug development in preclinical studies and subsequent clinical trials[1]–[5]. Many studies have attempted to predict drug properties, such as solubility and pKa[6]–[9], for drug repositioning[10]–[15] by applying various features and representation methods to the investigated molecules and proteins. Through machine learning, methods using various physicochemical properties[16]–[18], molecular fingerprint algorithms, such as the molecular access system (MACCS) [19] and Morgan[20] fingerprints[21]–[24], and graph neural networks[25]–[30] have been investigated. Additionally, protein therapeutics are widely used as drugs in almost all medical fields[31]. Therefore, numerous studies have been conducted on the prediction of protein properties using machine learning[32].

Nevertheless, unlike the various molecular fingerprint algorithms that have been developed[33], [34], protein fingerprint algorithms have rarely been studied. Several related studies have utilized protein fingerprints based on the three-dimensional structure of proteins[35], [36]. However, using these methods to obtain the three-dimensional structural information of proteins is expensive and time-consuming[37]. Unlike molecular fingerprints that use a simplified molecular-input line-entry system (SMILES), protein fingerprints cannot be simply analyzed using only amino acid sequences; thus, it is difficult to efficiently apply current protein fingerprints to machine learning.

Protein representation methods proposed in many previous studies have various limitations: (1) protein representation using structural information requires numerous resources for X-ray crystallography experiments[37], [38]; (2) sequence embedding: amino acids are simply treated as strings; hence, none of their properties are considered[39]–[41]; and (3) sequence computed features: in some cases, such as molecular weight, the same value is calculated for the same combination regardless of the sequence of amino acids; thus, different proteins are treated the same, and feature scaling is required according to the model[42].

In this study, we proposed a new method called amino acid molecular fingerprint repurposing-based protein (AmorProt) fingerprint. This algorithm identifies protein fingerprints by applying and repurposing various molecular fingerprint calculation algorithms to 20 amino acids. It has the advantages of fast calculation and utilization of the structure at the molecular level of amino acids. To verify the effectiveness of the AmorProt fingerprints, the following case studies were designed, as shown in **Figure S1, Supporting Information (SI)**. (1) Classification of amyloid-forming peptides: Pathologically, protein aggregation has been implicated in more than 40 human diseases, including Alzheimer's disease, Parkinson's disease, other neurodegenerative disorders, several cancers, and type II diabetes[43]–[46]. (2) Regression of the peptide isoelectric point (pI): The pI is an important physicochemical property that can be used to estimate the surface charge of a protein or peptide under various pH conditions. In particular, in therapeutic antibodies with generally low pIs, the tissue absorption is reduced, and the half-life is prolonged[47]–[49]. (3) Comparison of fingerprint- and feature-based methodologies: We compared the advantages and disadvantages of our fingerprint-based methodology with those of the easily applicable feature-based methodology. (4) Performance evaluation of tree-based machine learning algorithms and artificial neural networks: The machine learning and deep learning models were evaluated for the above two properties and two methodologies, and the versatility was verified by confirming the dependence of the dataset on each methodology. (5) Feature importance analysis. (6) Visualization of the protein space. Thus, the efficiency of the proposed AmorProt fingerprint can be verified when applied to machine learning fields, such as the prediction of protein properties.

**METHODS**

**Protein fingerprint algorithm**

AmorProt is a method for identifying protein fingerprints by repurposing and applying various molecular fingerprint calculation algorithms to 20 amino acids. In this study, as shown in **Figure 1**, the AmorProt fingerprint was calculated by repurposing the following molecular fingerprints commonly used for virtual screening or similarity searches[50]–[52]: MACCS fingerprint, extended-connectivity fingerprint 4 (ECFP4), extended-connectivity fingerprint 6 (ECFP6)[53], and RDKit fingerprint[54].

As shown in **Figure 1**, to obtain a protein fingerprint representing the molecular fingerprint of each amino acid of a sequence comprising 20 amino acids, the calculation method of the AmorProt fingerprint is as follows: (1) molecular fingerprints are sequentially calculated for each amino acid in the sequence of one protein or peptide and linked to one amino acid. (2) In the case of sequences with the same composition but different sequences or mutations, there must be differences in each protein fingerprint. Thus, each molecular fingerprint of amino acids is multiplied by a "smoothed" sine wave. (3) These corrected fingerprints are added in the column direction. (4) Finally, the values in each column are normalized to less than or equal to one by dividing each value by the maximum value.

In detail, the smoothed trigonometric function for the sine wave is used as follows:

$$T(p) = \frac{1}{W}\sin\frac{p}{A} + R$$

Here, $p$ is the position, $W$ is the wavelength, $A$ is the amplitude, and $R$ is the position of the displacement axis. In this study, $W$ and $A$ were set to 10 and $R$ to 0.85. Based on this, all molecular fingerprints are redefined by multiplying $T(p)$ for that position $p$. For example, the first amino acid has a molecular fingerprint of 0 or 0.86, and the second amino acid has a molecular fingerprint of 0 or 0.87. In this way, different protein fingerprints can be calculated for sequences having the same combination but different sequences. We came up with the idea of this calculation method from "positional encoding" in the attention mechanism[55].

**Protein feature-based methods**

For performance comparison with AmorProt fingerprints, we used PyBioMed[56], a versatile open-source library for various molecular representations of chemicals, proteins, and DNAs, as a case study for feature-based methodologies, such as various physicochemical properties of proteins. Using this, 9,920 features can be estimated based on the protein sequence; therefore, it is used in protein-related machine learning research[57], [58]. However, in this study, we eliminated 8,000 features, including tripeptide compositions, which are not calculated for some sequences, and only used 1,340 features that can be calculated for all sequences of the amyloid and pI datasets[56], as shown in **Figure 1(f)**.

**Amyloid-forming sequence database**

The WALTZ-DB 2.0[59], a database of experimentally determined amyloidogenic peptide sequences, was used in this study. This database comprises 229 peptide sequences experimentally validated using electron microscopy and thioflavin-T binding assay and 98 sequences collected from the literature. Moreover, this dataset comprised 1,416 hexapeptide sequences, including 515 amyloid and 901 non-amyloid sequences[59].

**Protein isoelectric point database**

We used the peptide dataset used in Isoelectric Point Calculator 2.0 (IPC 2.0)[60], a web server for the prediction of pI using machine learning. Peptides containing only 20 amino acids were used, which comprised a total of 119,085 amino acid sequences. Additionally, the experimental pI values obtained from high-resolution isoelectric focusing indicated that these peptides comprised 14.6 amino acids on average[60]. The pI value distribution of the dataset is shown in **Figure S2**, where the high and low groups can be seen at approximately 5.

**Model performance evaluation**

The light gradient boosting machine (LGBM) and multi-layer perceptron (MLP) models, widely used because of their versatility, were used to evaluate the performance of tree-based machine learning algorithms and deep learning neural networks, respectively[61]–[67]. The hyperparameters of the LGBM model were tuned through a grid search cross-validation process, and the MLP model was optimized through a manual hyperparameter tuning process. The final model architectures and hyperparameters of each MLP are shown in **Figures S5 and S6,** respectively.

As shown in **Figure S3**, 80 and 20% of both the amyloid and pI datasets were used as training and test sets, respectively. Subsequently, for the training set, five models were trained through five-fold cross-validation, and the final model performance was evaluated by predicting the ensemble of these five models in the test set. In the case of MLP, to strictly separate the test set from the model training process, feature scaling of the test set was conducted based on the training set alone.

**Comparison of dataset dependence**

In this case study, we compared the dependence of the training set on the feature scale for each methodology. In the case of LGBM, feature scaling may not have a significant effect on the performance of the model[61]; however, deep learning models, such as MLP, ideally receive input values in the [0, 1] range[68]. Therefore, the dependence of each methodology on the training set was compared only for the MLP model, wherein the feature- and fingerprint-based methods were compared.

To conduct a more rigorous case study, we identified the problems occurring when the features of the training set comprising restricted proteins are more limited in space than those of the test set comprising proteins outside the range for the 119,085 sequences used as the pI data, as presented in **Table S1**. We identified five features with the largest variance when each of the 1,340 features available in PyBioMed was calculated. In particular, the training sets accumulated based on limited experimental data are sensitive to these problems. Therefore, as shown in **Figure S4**, sequences with a value of 10 or less, based on each feature, were separated into the training set, and those with the maximum value (100) were separated into the test set. For each experiment, as shown in **Figure S3**, feature scaling was conducted based on the training set, assuming that the experimental value was known[69].

## RESULTS AND DISCUSSION

**Amyloid classification**

In this study, the binary classification of whether a specific peptide forms amyloid was compared using the AmorProt fingerprint-based and PyBioMed feature-based methods for the LGBM and MLP models. The detailed architecture, hyperparameters, and training process of the MLP are shown in **Figure S5**. Consequently, when the receiver operating characteristic area under the curve (ROC AUC), sensitivity (Sn), and specificity (Sp) were measured in both five-fold cross-validation and test set prediction, as presented in **Table S2** and shown in **Figure 2**, the results using the fingerprint-based method were better than those using the feature-based method in both models. In particular, in the test set prediction, the best results were obtained by LGBM using the AmorProt fingerprint expressed only as ECFP4, with an AUC value of 0.794. Additionally, for MLP, using the AmorProt fingerprint expressed only as ECFP4 provided the best results, with an AUC value of 0.760, compared to the feature-based method with an AUC value of 0.744. Overall, the 1024-dimensional fingerprint-based method was more effective than the 1340-dimensional feature-based method. Furthermore, when comparing models, as shown in **Figure S6,** the AUC values of MLP and LGBM were 0.760 and 0.794, respectively, indicating that LGBM was more effective than MLP. This implies that the AmorProt fingerprint is effective not only for artificial neural networks but also for tree-based machine learning algorithms, which are relatively simple models. Moreover, features such as the amino acid composition ratio are calculated as the same value, regardless of the amino acid sequence. Therefore, one limitation is that features are expressed as the same value for other sequences. However, the AmorProt fingerprint method solves these limitations using trigonometric functions, indicating that this method is applicable not only to proteins but also to short sequences comprising peptides, such as our amyloid dataset. Therefore, in this experiment, the fingerprint-based method was more effective than the feature-based one.

**Isoelectric point regression**

The pI regression of the peptide was predicted and compared between the two models and two methods. The detailed architecture, hyperparameters, and training process of the MLP are shown in **Figure S7**. As presented in **Table S3** and shown in **Figure 3**, in the case of LGBM, in the test set, the $R^2$ value was 0.971 when using the fingerprint-based method, whereas, in the case of the feature-based method, the $R^2$ value was slightly high, at 0.973. Additionally, in the case of the feature-based method, only 753-dimensional features were used; however, in the case of the fingerprint-based method, although 4263-bit fingerprints were used, the performance was not improved. In the case of MLP, in the test set, the fingerprint-based method expressed as 4263-bit fingerprints with an $R^2$ value of 0.961 was more effective than the feature-based method with an $R^2$ value of 0.935. In particular, different methods may be better depending on the two models; however, no significant difference was observed between their performances. These results suggest that using the fingerprint-based method in addition to the existing feature-based method is effective for developing protein property prediction models. Moreover, this study compares the two methods from another perspective related to the dependence of the dataset in the following case study.

**Comparison of dataset dependence**

In the case of MLP, to receive an input in the [0, 1] range, feature scaling is required according to the input representation, and at this time, the scaling of unknown test set features is required based on the training set features. Therefore, the use of the feature-based method is difficult in situations where the training set is limited. However, the fingerprint-based method consists of values in the [0, 1] range; thus, this limitation can be overcome. Therefore, if the model depends on the training set according to this input representation method, it is difficult to predict new data with only a limited data space.

For this purpose, based on the dataset used for pI regression, among the features of PyBioMed calculated from all sequences, we selected five cases with the highest variance for each feature: HydrophobicityD1025, SolventAccessibilityD2025, NormalizedVDWVD3025, PolarizabilityD3025, and PolarityD3001. In all these cases, the minimum value was 0, and the maximum value was 100; sequences with a corresponding feature value of ≤10 were defined as the training set, and those with the maximum value were defined as the test set. Additionally, in each case, the test set was scaled using the min-max scaling method based on each training set.

Consequently, as presented in **Tables 1** and **S4–8**, we confirmed that the fingerprint-based method is more effective than the feature-based method in predicting the performance of the test set in all cases, with an average $R^2$ difference of 0.319. In particular, in the case of HydrophobicityD1025, in the five-fold cross-validation using the training set, the feature-based method had a better $R^2$ value of 0.804; however, the $R^2$ value using the test set decreased to 0.782, whereas that of the feature-based method was the best at 0.801. For the other cases, the feature-based method was superior in both the training and test sets, unlike the pI regression results of the previous entire dataset because of the difference in the smaller training set space. Therefore, the AmorProt fingerprint enables the development of a deep learning model that is relatively independent of the training set and can effectively predict datasets in a significantly different space, highlighting the versatility of this method. This suggests that the dependence on the feature scale of the limited protein space, such as expensive experimental data, can be resolved. Furthermore, as shown in **Figure 3**, in the entire dataset, we verified that the feature-based method was superior to LGBM. Moreover, this also suggests that fingerprint-based methods are advantageous for developing more generalized models through the validation of limited datasets.

**Feature importance analysis**

As the performance of the AmorProt fingerprint and its effects were confirmed above, the feature importance between each bit position of the MACCS fingerprint was measured to confirm the explainability of the model performance. In particular, as the MACCS fingerprint is calculated according to the presence or absence of a specific SMILES arbitrary target specification (SMARTS) pattern in a molecule, the relationship between the importance at the corresponding bit position and the amino acid with the corresponding pattern can be analyzed. Therefore, the Shapley additive explanation (SHAP) method was applied to LGBM and MLP using the AmorProt fingerprint and implemented as a MACCS fingerprint in amyloid classification and pI regression. The SHAP method has been widely used in recent related studies owing to its explanatory power of feature importance[70]–[74].

For amyloid classification, as shown in **Figure 4**, the SMARTS pattern at position 149, "[C;H3,H4]," is the most significant for both LGBM and MLP models, and the amino acids with this pattern are isoleucine (I), leucine (L), and valine (V). Additionally, in skeletal muscle, significant amounts of valine and isoleucine are present in the β-sheet, and leucine is present in the α-helix, all of which are known to be involved in protein aggregation[75], [76]. Moreover, for pI regression, the SHAP values of the two models were of high importance in the top four, especially at positions 129, 136, 140, and 142. Consequently, this not only indicates that each bit position of the AmorProt fingerprint is critically involved in the prediction performance of machine learning models but also confirms the possibility of a causal relationship between amino acids containing specific molecular structural patterns and the properties of the protein, such as the pattern position 149 in amyloid classification.

In summary, the advantages of the AmorProt fingerprint were verified by comparing the performances of the fingerprint- and feature-based methods. As presented in **Table 2**, the AmorProt fingerprint has the following advantages: (1) feature generation is possible only with sequences. (2) Each amino acid uses molecular structural information rather than simple strings. (3) As the smoothed trigonometric function values are multiplied according to the order of amino acid sequences, sequences with the same composition but different orders are generated differently with different fingerprints. (4) It does not depend on the training set because it does not require feature scaling.

**Protein space visualization**

Data visualization is an essential tool for uncovering hidden patterns in high-dimensional data and interpreting these patterns[77]. Similarly, the analysis of the two-dimensional protein space using dimensionality reduction can visualize the distribution of protein properties and interpret hidden patterns[77]–[79]. Accordingly, each AmorProt fingerprint was calculated with the best configuration in the test set prediction results for the amyloid and pI datasets and then visualized after two-dimensional reduction using the T-distributed stochastic neighborhood embedding (t-SNE) algorithm[80]. Thus, the distribution according to the attributes of each coordinate in the two-dimensional space can be confirmed. Consequently, as shown in **Figure 5**, we confirmed that each coordinate for the two properties was distributed in positions that could be distinguished according to each value. For example, if a scholar requires non-amyloid sequences, this method can be applied to avoid amyloid sequences by sampling proteins located in the lower-left corner of the protein space shown in **Figure 5(a)**. This suggests that the AmorProt fingerprint can be applied to unsupervised learning as it can be classified according to its properties.

## CONCLUSIONS

The AmorProt fingerprint method proposed in this study is a protein sequence representation method that effectively utilizes molecular fingerprints corresponding to 20 amino acids. The utility of this method ranges from machine learning algorithms to artificial neural networks, and its performance has been verified through amyloid classification and pI regression experiments. Consequently, this is an atomic-level structure-based protein representation method that can be utilized only with sequence information. As this method was effective in all the above experiments, it can be used with various previous methods. Moreover, this fingerprinting method, with a value in the range of [0, 1], does not require the feature scaling step required by feature-based methods; thus, its advantage of not depending on the training set has been verified. Therefore, this representation method can be conveniently used in the field of protein identification based on artificial intelligence.

## ACKNOWLEDGMENT

This work was supported by the National Research Foundation of Korea (NRF) grant funded by the Korea government (MSIT) (No. 2022R1F1A1074339, No. 2022R1C1C1009387).

## CODE AVAILABILITY

The main source code used in this study is available on the GitHub page https://github.com/mhlee216/AmorProt.

## DATA AVAILABILITY

All data sets used to generate the reported results can be found in the respective studies: WALTZ-DB 2.0[59] for the amyloid-forming sequence database (http://waltzdb.switchlab.org/) and IPC 2.0[60] for the protein isoelectric point database (http://ipc2.mimuw.edu.pl/).

# APPENDIX A. SUPPORTING INFORMATION

SS

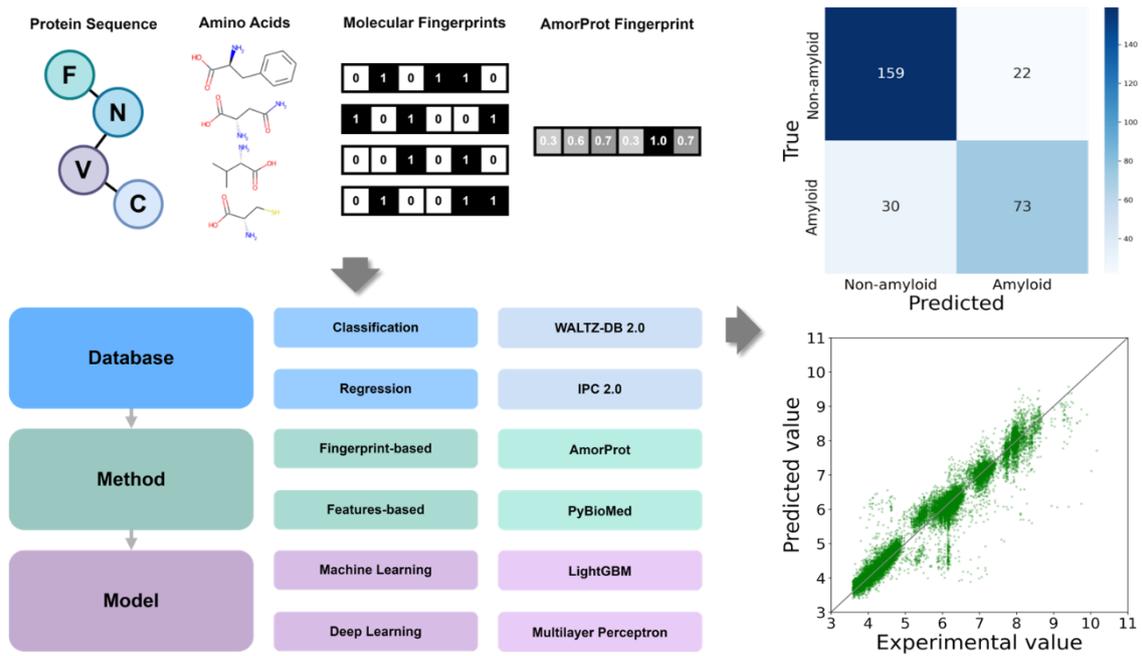

**Graphical Abstract**

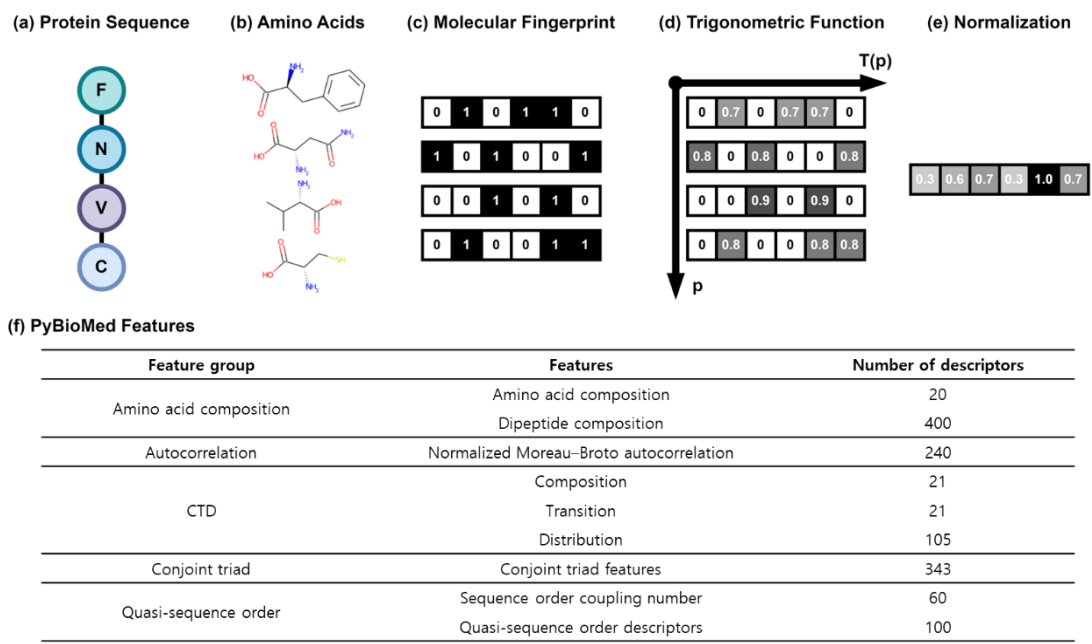

**Figure 1.** AmorProt fingerprint calculation method: **(a)** protein sequence. **(b)** Amino acids that constitute the sequence. **(c)** Calculation of the molecular fingerprints of each amino acid. **(d)** The result multiplied by the smoothed trigonometric function value $T(p)$ along each amino acid position $p$. **(e)** All values of each column are summed and then divided by the maximum value for normalization. **(f)** The features used by PyBioMed.

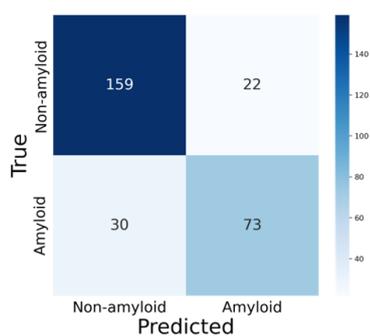 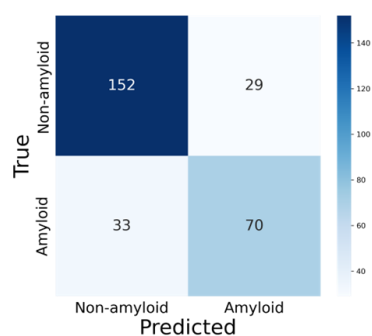

(a) LGBM  (b) MLP

(c)

| Method | Case | Dimension | LGBM | | | MLP | | |
|---|---|---|---|---|---|---|---|---|
| - | - | - | AUC | Sn | Sp | AUC | Sn | Sp |
| AmorProt | MACCS | 167 | 0.779 | 0.709 | 0.851 | 0.715 | 0.612 | 0.818 |
|  | ECFP4 | 1024 | **0.794*** | 0.709 | 0.878 | **0.760** | 0.680 | 0.840 |
|  | ECFP6 | 1024 | 0.781 | 0.689 | 0.873 | 0.733 | 0.631 | 0.834 |
|  | RDKit | 2048 | 0.767 | 0.650 | 0.884 | 0.757 | 0.680 | 0.834 |
|  | All fingerprints | 4263 | 0.789 | 0.689 | 0.890 | 0.748 | 0.650 | 0.845 |
| PyBioMed | All features | 1340 | 0.784 | 0.66 | 0.851 | 0.744 | 0.670 | 0.818 |
|  | Selected features | 753 | 0.784 | 0.66 | 0.856 | 0.724 | 0.641 | 0.807 |

**Figure 2.** Test set ensemble prediction results in the amyloid classification. **(a)** Prediction results of LGBM using the AmorProt fingerprint method comprising ECFP4. **(b)** Prediction result of MLP using the AmorProt fingerprint method comprising ECFP4. **(c)** Performance of all models. Bold letters indicate the best models, and * symbols represent the values used in **Figure 5**.

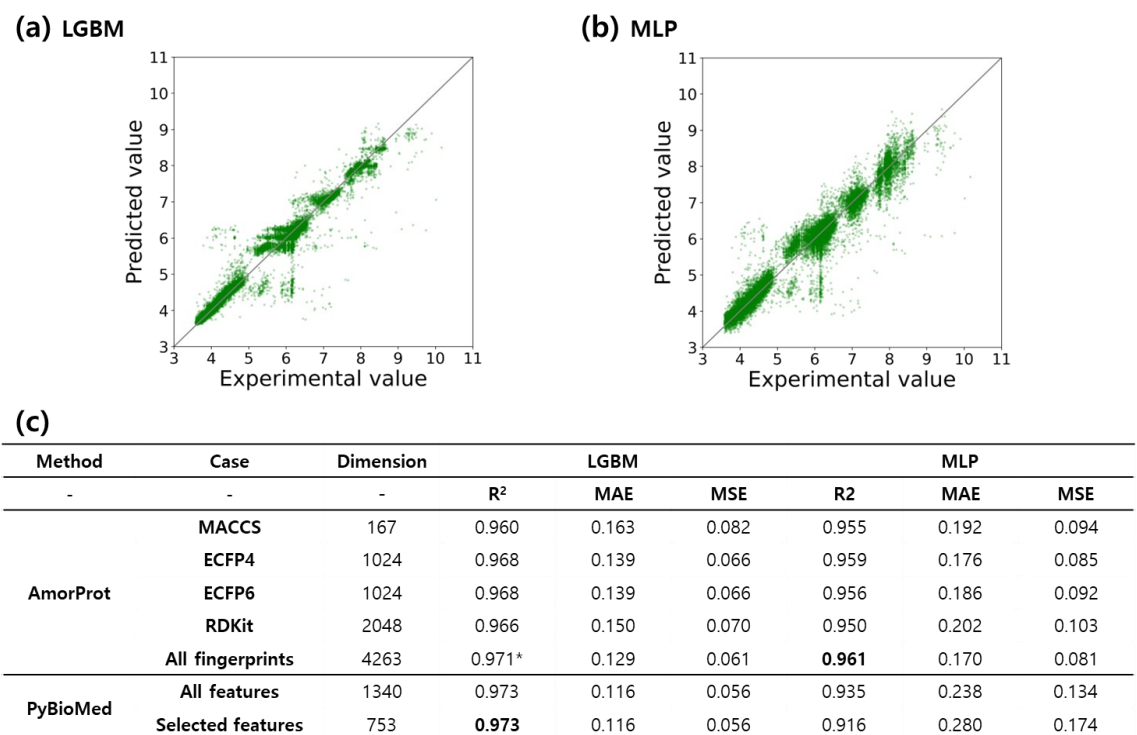

| Method | Case | Dimension | LGBM | | | MLP | | |
|---|---|---|---|---|---|---|---|---|
| - | - | - | R² | MAE | MSE | R2 | MAE | MSE |
| AmorProt | MACCS | 167 | 0.960 | 0.163 | 0.082 | 0.955 | 0.192 | 0.094 |
| | ECFP4 | 1024 | 0.968 | 0.139 | 0.066 | 0.959 | 0.176 | 0.085 |
| | ECFP6 | 1024 | 0.968 | 0.139 | 0.066 | 0.956 | 0.186 | 0.092 |
| | RDKit | 2048 | 0.966 | 0.150 | 0.070 | 0.950 | 0.202 | 0.103 |
| | All fingerprints | 4263 | 0.971* | 0.129 | 0.061 | **0.961** | 0.170 | 0.081 |
| PyBioMed | All features | 1340 | 0.973 | 0.116 | 0.056 | 0.935 | 0.238 | 0.134 |
| | Selected features | 753 | **0.973** | 0.116 | 0.056 | 0.916 | 0.280 | 0.174 |

**Figure 3.** Test set ensemble prediction results in pI regression. **(a)** Prediction results of LGBM using the AmorProt fingerprint method comprising all fingerprints. **(b)** Prediction result of MLP using the AmorProt fingerprint method comprising all fingerprints. **(c)** Performance of all models. Bold letters indicate the best models, and * symbols represent the values used in **Figure 5**.

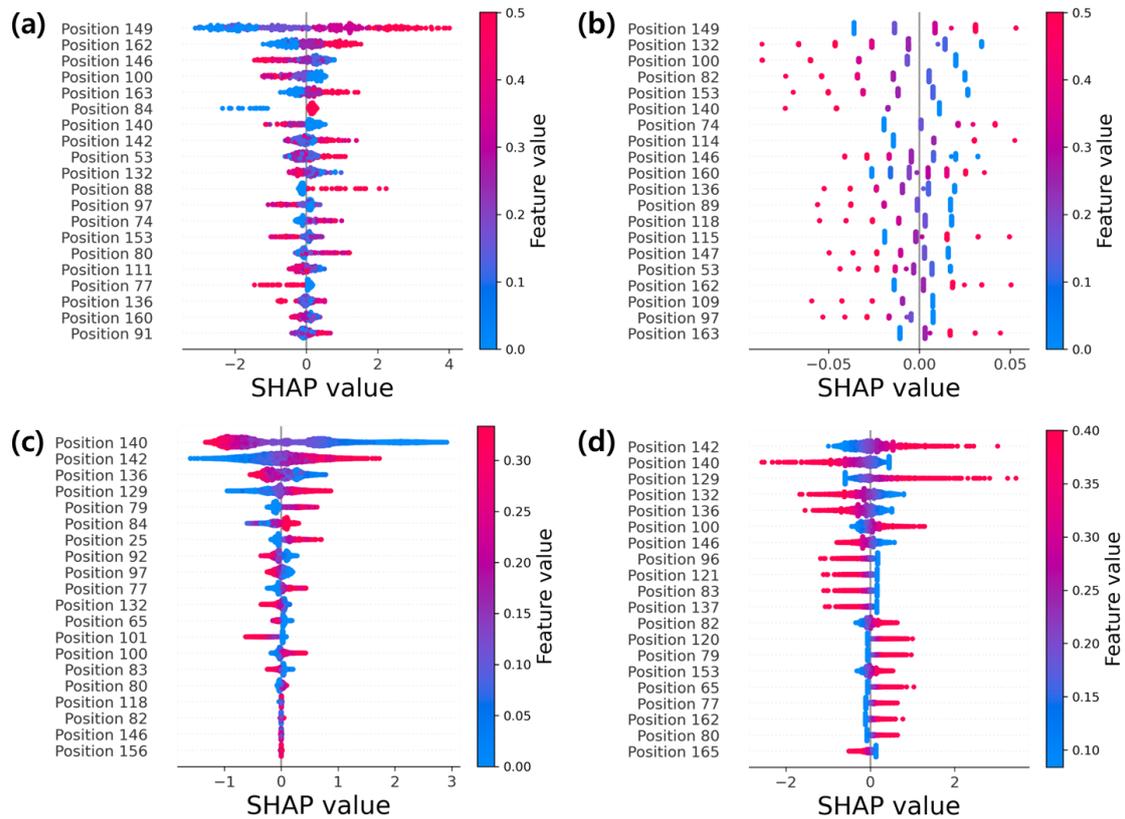

**Figure 4.** SHAP-based feature importance visualization. Results of **(a)** LGBM and **(b)** MLP for amyloid classification and those of **(c)** LGBM and **(d)** MLP for pI regression.

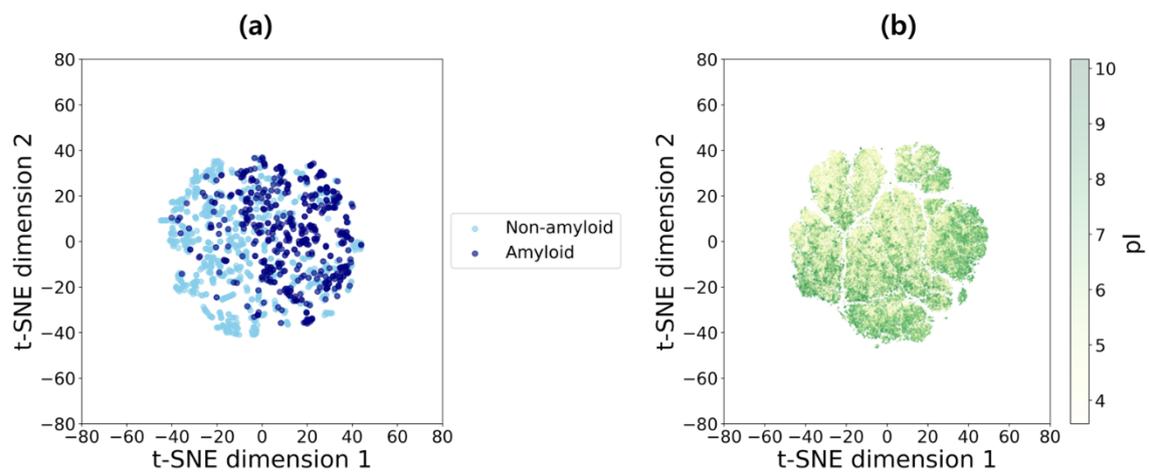

**Figure 5.** Protein space visualization based on t-SNE. **(a)** The amyloid and **(b)** pI datasets.

**Table 1.** Test set ensemble prediction results for each feature scaling case study in pI regression.

| Feature | Method | $R^2$ | MAE | MSE |
|---|---|---|---|---|
| HydrophobicityD1025 | AmorProt | **0.801** | 0.424 | 0.340 |
| | PyBioMed | 0.782 | 0.472 | 0.371 |
| SolventAccessibilityD2025 | AmorProt | **0.741** | 0.493 | 0.442 |
| | PyBioMed | 0.380 | 0.810 | 1.058 |
| NormalizedVDWVD3025 | AmorProt | **0.824** | 0.456 | 0.365 |
| | PyBioMed | 0.529 | 0.727 | 0.975 |
| PolarizabilityD3025 | AmorProt | **0.848** | 0.415 | 0.315 |
| | PyBioMed | 0.377 | 0.838 | 1.290 |
| PolarityD3001 | AmorProt | **0.789** | 0.485 | 0.398 |
| | PyBioMed | 0.339 | 0.966 | 1.245 |

**Table 2.** Disadvantages of previous protein expression methods and the advantages of the AmorProt fingerprint method.

| Prior Methods | Examples | Disadvantages of previous methods | Advantages of AmorProt Fingerprint |
|---|---|---|---|
| **Structure information** | • **Convolutional neural network**<br>• **Graph neural network** | • Requires a lot of time and money for X-ray experiments.<br>• Extensive computer resources are required. | • As structures of amino acids are used, the calculation is possible if only the sequence is present.<br>• It is scalable because various molecular fingerprint calculation methods are used for already-known structures of amino acids. |
| **Sequence embedding** | • **One-hot encoding**<br>• **Word embedding** | • Padding is computationally inefficient.<br>• Only representations dependent on the training set are possible.<br>• Amino acids are represented only as strings. | • A fixed dimension equal to the total length of the molecular fingerprints used.<br>• Calculated for each sequence regardless of the training set.<br>• Uses molecular fingerprints calculated using structures of amino acids not simple strings. |
| **Features calculated by sequence** | • **PyBioMed** | • Regardless of the order of the amino acid sequence, the same value can be calculated for the same combination.<br>• Certain models require feature scaling, which makes them dependent on the training set. | • As it is calculated using the smoothed trigonometric function, sequences with the same composition but different order are generated differently with different fingerprints.<br>• No scaling is required by normalization; hence, it does not depend on the training set. |